\documentclass[prd,aps,12pt]{revtex4}
\usepackage{epsfig}
\begin{document}

\title{Volume dependences from lattice chiral perturbation theory}

\author{Bu\=gra Borasoy}

\affiliation{Helmholtz-Institut f\"ur Strahlen- und Kernphysik (Theorie),
             Universit\"at Bonn, Nu{\ss}allee 14-16, D-53115 Bonn, Germany}

\author{Randy Lewis}

\affiliation{Department of Physics, University of Regina, Regina, SK,
S4S 0A2, Canada}

\begin{abstract}
The physics of pions within a finite volume is explored using
lattice regularized chiral perturbation theory. 
This regularization scheme permits a straightforward computational
approach to be used in place of analytical continuum techniques.
Using the pion mass, decay constant, form factor and charge radius as examples,
it is shown how numerical results for volume dependences are obtained at the
one-loop level from simple summations.
\end{abstract}

\maketitle

\section{Introduction}\label{sec:intro}

Lattice QCD is one of the key tools for studying hadronic physics.\cite{Wilson}
It is a numerical technique that employs a finite spatial volume, a finite
extent in Euclidean time, and a nonzero spacing between sites on the spacetime
lattice.  Lattice QCD practitioners also choose unphysically large masses
for
up and down quarks due to the extreme cost of simulations at their physical
values.

The extrapolation to physical up and down quark masses can in principle be
performed
by using the low energy effective theory for continuum QCD, called
chiral perturbation theory (ChPT).\cite{GL8485}
The Lagrangian of ChPT contains an
infinite number of terms, but to a specific order in the small chiral expansion
parameters (for the pure pion theory these are $m_\pi^2/(4\pi f_\pi)^2$ and
$p^2/(4\pi f_\pi)^2$ with $p$ being a small four-momentum) the number of terms
is finite.  ChPT has established
itself as a valuable formalism for hadronic physics, and its use in
connection to lattice QCD is just one important example.

Similarly, the extrapolation in lattice spacing can be discussed within the
effective theory for {\em lattice} QCD, which is simply ChPT extended to
include the effects of the nonzero lattice spacing, $a$.
This requires the addition of an infinite number of new terms to the continuum
ChPT Lagrangian,
each of which is proportional to some positive power of $a$.  To a
specific
order in the lattice spacing expansion, the number of $a$-dependent terms is
finite and the numerical values of their coefficients can be determined in
principle by matching to a particular definition of lattice QCD.
Different lattice QCD Lagrangians (Wilson, Symanzik-improved, etc.)
correspond to different lattice ChPT
coefficients for the $a$-dependent counterterms.
All of these additional terms become irrelevant in the continuum limit.

Lattice ChPT can be defined within a continuum quantum field theory formalism,
using (for example) dimensional regularization to handle ultraviolet
divergences and retaining the lattice spacing only as prefactor for the
$a$-dependent Langrangian counterterms mentioned above.\cite{Sharpe,Bar}
Another option is to define lattice ChPT in an explicitly discrete
spacetime.\cite{Rebbi,ShuSmi,LO,BLO}
The lattice spacing now plays the role of ultraviolet regulator in addition
to being the expansion parameter for the $a$-dependent Lagrangian counterterms.
With this approach, the lattice spacing appears explicitly
in propagators and vertices and also in limits of integration for Feynman
loop diagrams.
The continuum and discrete methods are essentially equivalent when
the inverse lattice spacing lies beyond the regime of ChPT
($1/a>\Lambda_\chi\sim m_\rho\sim4\pi f_\pi$) as is the case in typical
lattice QCD simulations.
One method or the other may be preferred for ease of use, or for theoretical
discussions of the convergence properties of the ChPT expansion.\cite{LO,BLO}

The extrapolation in lattice volume within the framework of ChPT requires in 
general the inclusion of boundary-valued counterterms to the Lagrangian
due to explicit boundary conditions except (as shown by Gasser and
Leutwyler\cite{GLvolume1,GLvolume2}) for toroidal spacetime. In this case,
the only effect of finite volume is the straightforward
conversion of loop momentum integrals to loop momentum summations.
For a review of recent finite volume ChPT calculations,
see Ref.~\cite{Colangelo}.  Some of the latest studies in the pion sector
are those of Refs.~\cite{ColDur,ColHae,Linetal}.

In the present work, we explore the use of lattice regularized ChPT for
computing volume dependences.  The continuum limit must be identical to
any viable continuum regulator, but lattice regularization has the feature
of being easy to manage numerically.
Beginning from a Lagrangian that displays the lattice spacing explicitly
and also maintains exact chiral symmetry,\cite{LO}
one can simply derive the Feynman
propagators and vertices then type those directly into a computer program.
Loop diagrams are just summations of a finite number of momentum values and
the numerics are finite at every step.
For a sufficiently small lattice spacing, observables must be independent of
$a$.

A brief preliminary discussion of this work can be found in Ref.~\cite{BLM},
but a more detailed study is presented below.
Notation for the lattice regularized ChPT Lagrangian is established in
Sec.~\ref{sec:Lagrangian}.
The computational method is introduced in Sec.~\ref{sec:mass} by examining
the two-point pion
correlator.  This gives the volume dependence of the pion mass, which
reproduces a result already known from continuum
methods.\cite{GLvolume1,ColDur}  The two-point
correlator also gives an explicit expression for wave function renormalization
in the lattice regularized theory.
Section \ref{sec:fpi} contains a computation of volume effects on the
pion decay constant which agrees with published continuum
calculations.\cite{GLvolume1,ColHae}
New results are presented in Sec.~\ref{sec:pionff}: volume dependences of the
pion form factor and the pion charge radius.
Section \ref{sec:outlook} mentions some of the challenges that remain to be
addressed if lattice regularized ChPT is to be employed for the determination
of volume dependences beyond the one-loop level.
Appendix~\ref{feynmanrules} provides an explicit example of calculating
Feynman rules from the lattice ChPT action, and Appendix \ref{continuumlimit}
demonstrates the exact analytic agreement between volume dependences in
dimensional regularization
and in the continuum limit of lattice regularization.

\section{A discretized SU(2) chiral Lagrangian}\label{sec:Lagrangian}

The Lagrangian to be used in this work is an SU(2) version of the
SU(3) meson Lagrangian introduced in Ref.~\cite{LO}.
Although only a few terms are presently required,
here is the complete Lagrangian:
\begin{eqnarray}
{\cal L} &=& {\cal L}_2 + {\cal L}_4, \label{L} \\
{\cal L}_2 &=& \frac{f^2}{4}\left<\nabla^{(+)}_\mu U^\dagger\nabla^{(+)}_\mu
               U\right>
             - \frac{f^2}{4}\left<\chi^\dagger U+\chi U^\dagger\right>, \\
{\cal L}_4 &=& -\frac{1}{4}l_1\left<\nabla^{(\pm)}_\mu
                              U^\dagger\nabla^{(\pm)}_\mu U\right>^2
               -\frac{1}{4}l_2\left<\nabla^{(\pm)}_\mu
                              U^\dagger\nabla^{(\pm)}_\nu U\right>
                              \left<\nabla^{(\pm)}_\mu
                              U^\dagger\nabla^{(\pm)}_\nu U\right>
                              \nonumber \\
         &&  -\frac{1}{16}(l_3+l_4)\left<\chi^\dagger U+\chi U^\dagger\right>^2
             +\frac{1}{8}l_4\left<\nabla^{(\pm)}_\mu
                            U^\dagger\nabla^{(\pm)}_\mu U\right>
                            \left<\chi^\dagger U+\chi U^\dagger\right>
             -l_5\left<F_{\mu\nu}^LU^\dagger F_{\mu\nu}^RU\right>
             \nonumber \\
         &&  -\frac{i}{2}l_6\left<F_{\mu\nu}^L\nabla^{(\pm)}_\mu
                            U^\dagger\nabla^{(\pm)}_\nu U
                           +F_{\mu\nu}^R\nabla^{(\pm)}_\mu
                            U\nabla^{(\pm)}_\nu U^\dagger\right>
             +\frac{1}{16}l_7\left<\chi^\dagger U-\chi U^\dagger\right>^2
             \nonumber \\
         && -\frac{1}{4}(h_1+h_3-l_4)\left<\chi^\dagger\chi\right>
            +\left(2h_2+\frac{l_5}{2}\right)\left<F_{\mu\nu}^LF_{\mu\nu}^L
            +F_{\mu\nu}^RF_{\mu\nu}^R\right> \nonumber \\
         && -\frac{1}{16}(h_1-h_3-l_4)\left(
                         \left<\chi^\dagger U+\chi U^\dagger\right>^2
                        +\left<\chi^\dagger U-\chi U^\dagger\right>^2
                       -2\left<\chi^\dagger U\chi^\dagger U+U^\dagger\chi
                         U^\dagger\chi\right>\right), \label{L4}
\end{eqnarray}
where $\left<\ldots\right>$ denotes a trace, and summations over repeated
Lorentz indices $\mu$ and $\nu$ are understood.
$\chi$ is essentially the quark mass matrix,
\begin{equation}
\chi = 2B\left(\begin{array}{cc} m_u & 0 \\ 0 & m_d \end{array}\right).
\end{equation}
Throughout this work, we restrict ourselves to the isospin limit
$m_u=m_d\equiv m_q$.
We also choose the exponential representation for pions,
\begin{equation}
U(x) = \exp\left[\frac{i\tau^a\pi^a(x)}{f}\right],
\end{equation}
where $\tau^a$ is a Pauli matrix.
The external fields are
\begin{eqnarray}
L_\mu(x) &=& \exp\left[-ia\ell_\mu(x)\right]
           = \exp\left[-ia(V_\mu(x)-A_\mu(x))\right], \\
R_\mu(x) &=& \exp\left[-iar_\mu(x)\right]
           = \exp\left[-ia(V_\mu(x)+A_\mu(x))\right],
\end{eqnarray}
and the corresponding field strength tensors are discretized as follows:
\begin{eqnarray}
4ia^2F_{\mu\nu}^X &=& 4
          - X_\mu(x)X_\nu(x+a_\mu)X^\dagger_\mu(x+a_\nu)X^\dagger_\nu(x)
          \nonumber \\
  && - X_\nu(x)X^\dagger_\mu(x-a_\mu+a_\nu)X^\dagger_\nu(x-a_\mu)X_\mu(x-a_\mu)
     \nonumber \\
  && - X^\dagger_\mu(x-a_\mu)X^\dagger_\nu(x-a_\mu-a_\nu)X_\mu(x-a_\mu-a_\nu)
     X_\nu(x-a_\nu) \nonumber \\
  && - X^\dagger_\nu(x-a_\nu)X_\mu(x-a_\nu)X_\nu(x+a_\mu-a_\nu)X^\dagger_\mu(x)
\end{eqnarray}
where $X=L,R$.
As discussed in Ref.~\cite{LO}, a convenient way to avoid unphysical poles
in the spectrum while maintaining invariance under parity is to use a
nearest-neighbour derivative in the leading order Lagrangian,
\begin{equation}
\nabla^{(+)}_\mu U(x) = \frac{1}{a}\left[R_\mu(x)U(x+a_\mu)L_\mu^\dagger(x)
                      - U(x)\right],
\end{equation}
and a symmetrized derivative at next-to-leading order,
\begin{equation}
\nabla^{(\pm)}_\mu U(x) = \frac{1}{2a}\left[R_\mu(x)U(x+a_\mu)L_\mu^\dagger(x)
                      - R_\mu^\dagger(x-a_\mu)U(x-a_\mu)L_\mu(x-a_\mu)\right].
\end{equation}
Notice that the Lagrangian in Eqs.~(\ref{L}-\ref{L4}) contains exactly the
same number of terms as the continuum SU(2) ChPT Lagrangian\cite{GL8485}.
As discussed in Sec.~\ref{sec:intro}, the most general ChPT Lagrangian would
contain
extra terms proportional to positive powers of the lattice spacing.  Since
we are presently interested in volume dependences at the continuum limit,
these extra terms are irrelevant and hence omitted for simplicity.

To conclude this section, we recall that the ChPT action is
\begin{equation}
S = a^4\sum_x{\cal L}(x) - \frac{1}{2}\sum_x\left<\ln\left[\frac{2
    (1-\cos\Phi(x))}{\Phi^2(x)}\right]\right>,
\end{equation}
where the second term is due to the integration measure.\cite{LO}
For SU(2),
\begin{equation}
\Phi(x) = \frac{-2i}{f}\left(\begin{array}{ccc} 0 & \pi^3(x) & -\pi^2(x) \\
                                                -\pi^3(x) & 0 & \pi^1(x) \\
                                                \pi^2(x) & -\pi^1(x) & 0
                                                \end{array}\right).
\end{equation}

\section{The pion mass and wave function renormalization}\label{sec:mass}

\begin{figure}[tb]
\includegraphics[width=15cm]{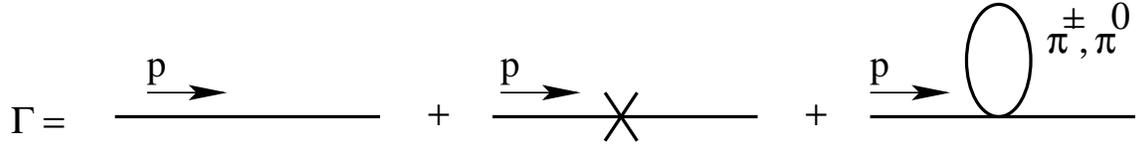}
\caption{Feynman diagrams contributing to the pion two-point correlator
         at one-loop level in ChPT.}
\label{fig:massfeyn}
\end{figure}

The Feynman diagrams for the one-loop pion two-point correlator are shown in
Fig.~\ref{fig:massfeyn}.  To evaluate them within lattice regularization, we
choose a hyper-rectangular lattice with lattice spacing $a$ in all four
spacetime directions.  The lattice is chosen to have $N_s$ sites in each of
the spatial directions and $N_t$ sites in the temporal direction.  Our goal
is to consider the dependence of observables on spatial volume in the double
limit $a\to0$, $aN_t\to\infty$ with $aN_s$ held fixed.

With Feynman vertices obtained from the Lagrangian of
Sec.~\ref{sec:Lagrangian} (see Appendix~\ref{feynmanrules} for the
derivations),
the three diagrams of Fig.~\ref{fig:massfeyn} respectively become
\begin{eqnarray}
\Gamma_{\rm LO} &=& -x_\pi^2 - \frac{2}{a^2}\sum_\mu(1-\cos ap_\mu),
            \label{massLO} \\
\Gamma_{\rm NLO}^{(a)} &=& -\frac{2}{3a^4f^2} - \frac{2x_\pi^4}{f^2}(l_3+l_4)
                           - \frac{2l_4x_\pi^2}{a^2f^2}\sum_\mu\sin^2ap_\mu, \\
\Gamma_{\rm NLO}^{(b)} &=& \frac{1}{6N_s^3N_ta^4f^2}\sum_k
           \bigg(112+5a^2x_\pi^2
              -20\sum_\mu\cos ap_\mu-20\sum_\mu\cos ak_\mu \nonumber \\
           && +12\sum_\mu\cos ap_\mu\cos ak_\mu\bigg)D(k), \label{massNLO}
\end{eqnarray}
where
\begin{equation}\label{piprop}
D(k) = \frac{1}{a^2x_\pi^2+2\sum_\mu(1-\cos ak_\mu)}
\end{equation}
is the pion propagator and
\begin{equation}
x_\pi = \sqrt{2Bm_q}
\end{equation}
is the lowest-order pion mass in the continuum limit.
The symbol ``$\sum_k$'' in Eq.~(\ref{massNLO}) represents a sum
over available lattice 4-momenta; for any function $F$, this means
\begin{equation}
\sum_kF(k_1,k_2,k_3,k_4) \equiv
 \sum_{n_1=1}^{N_s}\sum_{n_2=1}^{N_s}\sum_{n_3=1}^{N_s}\sum_{n_4=1}^{N_t}
 F\left(\frac{2\pi n_1}{aN_s},\frac{2\pi n_2}{aN_s},\frac{2\pi n_3}{aN_s},
 \frac{2\pi n_4}{aN_t}\right).
\end{equation}
Notice that the middle diagram in Fig.~\ref{fig:massfeyn} includes the
measure contribution as well as the tree-level ${\cal L}_4$ contributions.

The pion mass is defined as the energy of a stationary pion.  The corresponding
expression for the pion mass is the value of $ip_4$ which solves $\Gamma=0$
when $\vec p=\vec 0$, where $\Gamma$ is the sum of the three diagrams
\begin{equation}
\Gamma = \Gamma_{\rm LO} + \Gamma_{\rm NLO}^{(a)} + \Gamma_{\rm NLO}^{(b)}.
    \label{Gammaeqn}
\end{equation}
The result is
\begin{eqnarray}
M_\pi &=& \frac{2}{a}{\rm arcsinh}\left(\frac{aX_\pi}{2}\right),
          \label{masseqn1} \\
X_\pi^2 &=& x_\pi^2 + \frac{2x_\pi^4}{f^2}l_3
           +x_\pi^2\sum_k\frac{(3-2\cos ak_4)}{2N_s^3N_ta^2f^2}D(k) + O(a).
\label{masseqn2}
\end{eqnarray}
Given numerical values for the Lagrangian parameters $f$, $Bm_q$ and $l_3$,
the pion mass can now be computed directly from Eqs.~(\ref{masseqn1}) and
(\ref{masseqn2}) for any lattice spacing and volume.  As $a\to0$ the loop
diagram diverges and these divergences are cancelled by the $a$ dependence
of the bare Lagrangian parameters $f$, $Bm_q$ and $l_3$.
(Since dimensional regularization retains no power divergences, $f$ and
$Bm_q$ would be scale invariant in that scheme.  Lattice regularization
does retain power divergences as $a\to0$ so the parameters $f$ and $Bm_q$
do have $a$ dependence in this scheme.)
For any $a\neq0$ the loop diagram is finite, and for sufficiently small
$a$ the renormalized pion mass is independent of lattice spacing.

To extract the volume dependence of the pion mass, one needs only the
{\em difference} of $M_\pi$ at two different spatial volumes.
The first two diagrams in Fig.~\ref{fig:massfeyn} cancel in this difference
leaving only the loop diagram.
As $a\to0$, the difference between two volumes must be finite because the
only available Lagrangian counterterms were in the first two diagrams.
The quantities $x_\pi$ and $f$ appearing in the loop diagram are the
leading chiral-order expressions for the mass and decay constant in the
continuum limit.  Following Ref.~\cite{ColDur}, we employ $f=86.2$ MeV.
One would expect results to become independent of lattice spacing for
$a\lesssim1/(4\pi f_\pi)\sim0.2$ fm, and we will choose $N_t\gg N_s$ so that the
temporal direction will not affect our extraction of spatial volume effects
in any significant way.

\begin{figure}[tb]
\includegraphics[width=10cm]{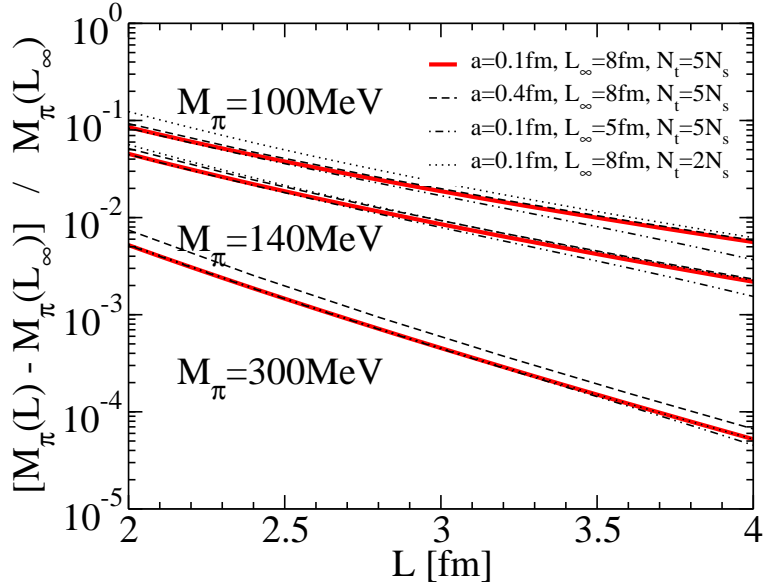}
\caption{Fractional change in the pion mass as a function of spatial volume.}
\label{fig:massplot}
\end{figure}

Figure \ref{fig:massplot} displays numerical results for the fractional
change in the pion mass as a function of spatial volume, relative to the
infinite volume pion mass, for $M_\pi(L_\infty)=100$, 140 and 300 MeV,
corresponding to $x_\pi=100$, 142 and 321 MeV respectively.
The computation at ``infinite'' volume, $L_\infty^3$, is performed numerically
simply by choosing a volume large enough to offer negligible deviations if the
volume is increased yet further.  Figure \ref{fig:massplot} shows explicitly
the dependence of numerical results on changes to $a$, $L_\infty$, and $N_t$.
As expected, heavier pions have an increased sensitivity to lattice spacing
because loop integrals depend on the product $ax_\pi$.
The computation at $a=0.1$ fm, $L_\infty=8$ fm and $N_t=5N_s$ produces
a fractional volume dependence for the pion mass that agrees with the known
continuum result\cite{GLvolume1,ColDur} to within the resolution of this plot
for the full range shown, $2~{\rm fm}<L<4$ fm.
As a confirming cross-check, this known continuum result is derived
analytically from our lattice
regularized expression in Appendix~\ref{continuumlimit}.

In addition to the pion mass, Eq.~(\ref{Gammaeqn}) also leads to an expression
for the wave function renormalization factor that will be required for all of
the observables to be addressed below.
Up to irrelevant lattice spacing effects, the two-point correlator can be
parametrized as
\begin{eqnarray}
\Gamma &=& -\left[\tilde p^2+x_\pi^2+\Sigma(-\tilde p^2)\right] \\
       &=& -\left[\tilde p^2+x_\pi^2+\Sigma(X_\pi^2)+(-\tilde p^2-X_\pi^2)
           \Sigma^\prime(X_\pi^2)+\delta\Sigma(-\tilde p^2)\right]
\end{eqnarray}
where
\begin{equation}
\tilde p^2 = \frac{4}{a^2}\sum_\mu\sin^2\left(\frac{ap_\mu}{2}\right),
\end{equation}
and $\delta\Sigma(-\tilde p^2)$ vanishes at least as quickly as
$(-\tilde p^2-X_\pi^2)^2$ for $-\tilde p^2\to X_\pi^2$.
The renormalization factor, $Z$, is defined by
\begin{equation}
\Gamma = -\left[\frac{\tilde p^2+X_\pi^2}{Z}+\delta\Sigma(-\tilde p^2)\right],
\end{equation}
since $X_\pi^2= x_\pi^2+\Sigma(X_\pi^2)$
which leads to
\begin{equation}
Z \equiv \frac{1}{1-\Sigma^\prime(X_\pi^2)}.
\end{equation}
One can read $\Sigma(-\tilde p^2)$ directly from
Eqs.~(\ref{massLO}-\ref{massNLO}) by choosing $\vec p=\vec 0$, and this gives
\begin{equation}\label{Z}
Z = 1 - \frac{2x_\pi^2l_4}{f^2} + \frac{1}{3N_s^3N_ta^2f^2}\sum_k(5-3\cos ak_4)
    D(k) + O(a).
\end{equation}

\section{The pion decay constant}\label{sec:fpi}

\begin{figure}[tb]
\includegraphics[width=15cm]{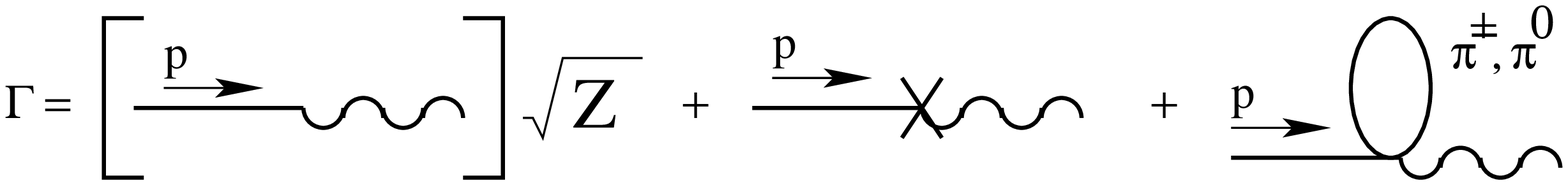}
\caption{Feynman diagrams contributing to the pion decay constant
         at one-loop level in ChPT.  A wavy line denotes an axial vector
         current insertion.}
\label{fig:fpifeyn}
\end{figure}

The three Feynman diagrams of Fig.~\ref{fig:fpifeyn} represent the three
contributions to the pion decay constant up to one-loop order.
Using the vertices and propagators from the Lagrangian in
Eqs.~(\ref{L}-\ref{L4}), one finds the following expressions for
those three diagrams,
\begin{eqnarray}
G_{\rm LO}\sqrt{Z} &=& \frac{i\sqrt{2}}{a}f\left[\sin ap_\mu+2i\sin^2\left(
               \frac{ap_\mu}{2}\right)\right]\sqrt{Z}, \\
G_{\rm NLO}^{(a)} &=& \frac{i2\sqrt{2}}{af}x_\pi^2l_4\exp\left(\frac{iap_\mu}
                      {2}\right)\cos \left( \frac{ap_\mu}{2} \right) \sin(ap_\mu), \\
G_{\rm NLO}^{(b)} &=& -i\left[\sin ap_\mu+2i\sin^2\left(\frac{ap_\mu}{2}\right)
                      \right]\frac{\sqrt{2}}{3N_s^3N_ta^3f}\sum_k(7
                      -3\cos ak_\mu)D(k).
\end{eqnarray}
Choosing a stationary pion ($\vec p=\vec 0$) and inserting Eq.~(\ref{Z}) for
the wave function renormalization factor leads to
\begin{equation}
G_{\rm LO} + G_{\rm NLO}^{(a)} + G_{\rm NLO}^{(b)}
= \frac{i\sqrt{2}}{a}f_\pi\left[\sin ap_4+O(a)\right]
\end{equation}
where the one-loop pion decay constant is
\begin{equation}\label{F}
f_\pi = f + \frac{x_\pi^2l_4}{f} - \frac{1}{2N_s^3N_ta^2f}\sum_k(3-\cos ak_4)
        D(k) + O(a).
\end{equation}

\begin{figure}[tb]
\includegraphics[width=10cm]{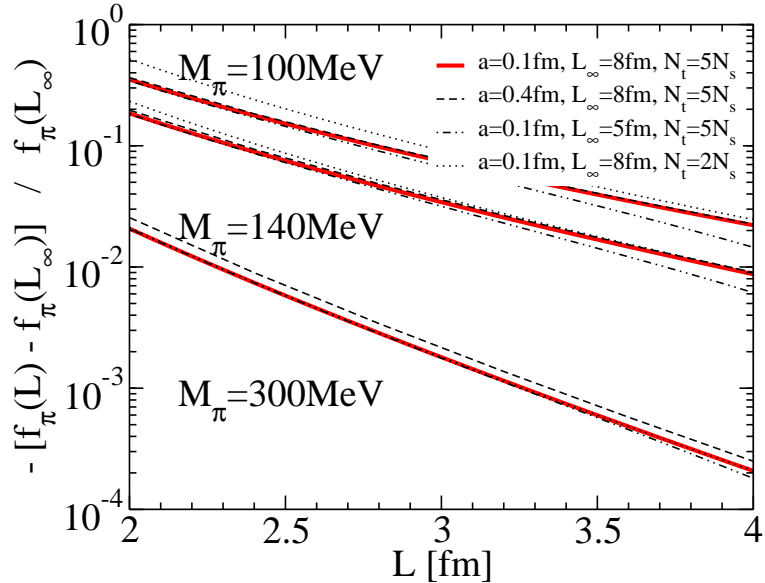}
\caption{Fractional change in the pion decay constant as a function of spatial
         volume.}
\label{fig:fpiplot}
\end{figure}

In the difference between $f_\pi$ computed from two different lattice volumes,
the first two terms in Eq.~(\ref{F}) subtract away.  To this chiral order,
the remaining parameters $x_\pi$ and $f$ can be set to the (infinite volume)
physical pion mass and decay constant.  The resulting volume dependence of
the pion decay constant is displayed in Fig.~\ref{fig:fpiplot}.
The magnitude of the volume dependence is similar to that of the pion mass
plotted in Fig.~\ref{fig:massplot}, but the sign differs --- the decay constant
is reduced as the volume shrinks, whereas the mass grows with shrinking volume.
The computation at $a=0.1$ fm, $L_\infty=8$ fm and $N_t=5N_s$ produces
a fractional volume dependence for the pion decay constant that agrees with
the known
continuum result\cite{GLvolume1,ColHae} to within the resolution of this plot
for the full range shown, $2~{\rm fm}<L<4$ fm.

The careful reader will notice that Fig.~\ref{fig:fpiplot} has a slightly
different normalization from the corresponding plot in Ref.~\cite{BLM};
this difference is higher order in the ChPT expansion, and is due to use
of the physical mass and decay constant in Ref.~\cite{BLM} in place of
the lowest-order parameters $x_\pi$ and $f$.
Figures \ref{fig:massplot} and \ref{fig:fpiplot}
of the present work show the familiar ratio,
$[f_\pi(L)/f_\pi(L_\infty)-1]/[M_\pi(L)/M_\pi(L_\infty)-1] = -4$,
expected from Ref.~\cite{GLvolume1}.

\section{The pion form factor and charge radius}\label{sec:pionff}

\begin{figure}[tb]
\includegraphics[width=15cm]{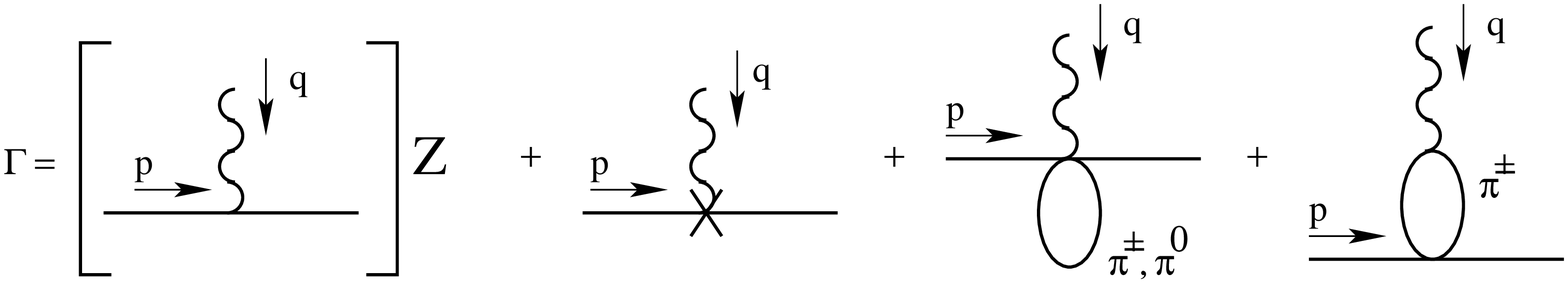}
\caption{Feynman diagrams contributing to the pion electromagnetic form factor
         at one-loop level in ChPT.  A wavy line denotes a photon.}
\label{fig:pionfffeyn}
\end{figure}

The pion electromagnetic form factor is obtained from the Feynman diagrams of
Fig.~\ref{fig:pionfffeyn}, and the charge radius can be extracted from the
slope of the form factor at vanishing photon 4-momentum.
Using the Lagrangian of Eqs.~(\ref{L}-\ref{L4}), the four diagrams evaluate as
follows,
\begin{eqnarray}
H_{\rm LO} &=& \frac{2 Z}{a}\exp\left(\frac{-iaq_\mu}{2}\right)\sin
               a\left(\frac{p+p^\prime}{2}\right)_\mu \\
H_{\rm NLO}^{(a)} &=& \frac{2l_4x_\pi^2}{af^2}\exp\left(\frac{-iaq_\mu}{2}
                      \right)\cos\left(\frac{aq_\mu}{2}\right)\sin
                      a(p+p^\prime)_\mu \nonumber \\
                  &&+ \frac{2l_6}{a^3f^2}\exp\left(\frac{-iaq_\mu}{2}\right)
                      \sum_\nu\left[\sin ap_\mu\sin ap^\prime_\nu-\sin
                      ap^\prime_\mu\sin ap_\nu\right]\sin aq_\nu\cos\left(
                      \frac{aq_\mu}{2}\right), \\
H_{\rm NLO}^{(b)} &=& \frac{-10}{3N_s^3N_ta^3f^2}\exp\left(\frac{-iaq_\mu}{2}
                      \right)\sum_k\sin\left(\frac{a(p+p^\prime)_\mu}{2}\right)
                      D(k), \\
H_{\rm NLO}^{(c)} &=& \frac{4}{N_s^3N_ta^3f^2}\exp\left(\frac{-iaq_\mu}{2}
                      \right)\sum_k\sin a(k+q/2)_\mu\sum_\nu\cos a(p-k)_\nu
                      D(k)D(k+q),
\end{eqnarray}
where $p$ is the incoming pion momentum, $q$ and $\mu$ are the incoming
momentum and Lorentz index of the external photon, and $p^\prime\equiv p+q$.
The contribution from $H_{\rm NLO}^{(c)}$ can be simplified by removing
terms that are odd under interchange of $k$ and $-(k+q)$, since these
vanish after summation over $k$.  The result is
\begin{equation}
H_{\rm NLO}^{(c)} = \frac{4}{N_s^3N_ta^3f^2}\exp\left(\frac{-iaq_\mu}{2}
                    \right)\sin a\left(\frac{p+p^\prime}{2}\right)_\mu
                    \sum_k\sin^2a(k+q/2)_\mu D(k)D(k+q).
\end{equation}
The pion form factor, $F(q^2)$, can be obtained explicitly by choosing
$\mu=4$ as follows,
\begin{equation}
H_{\rm LO} + H_{\rm NLO}^{(a)} + H_{\rm NLO}^{(b)} + H_{\rm NLO}^{(c)}
= \frac{2}{a}F(q^2)\exp\left(\frac{-iaq_4}{2}\right)\sin
  a\left(\frac{p+p^\prime}{2}\right)_4
\end{equation}
which gives
\begin{eqnarray}
F(q^2) &=& 1 + \frac{l_6}{a^2f^2}
               \sum_\nu\left[\sin ap_4\sin ap^\prime_\nu-\sin
               ap^\prime_4\sin ap_\nu\right]\sin aq_\nu\cos\left(
               \frac{aq_4}{2}\right) \nonumber \\
          && - \frac{1}{N_s^3N_ta^2f^2}\sum_k\cos ak_4D(k)
             + \frac{2}{N_s^3N_ta^2f^2}\sum_k\sin^2 a\left(k+\frac{q}{2}
               \right)_4 D(k)D(k+q) \nonumber \\
          && + O(a). \label{pionff}
\end{eqnarray}
It is interesting to consider the $q\to0$ limit, since vector current
conservation should require $F(0)=1$.  From Eq.~(\ref{pionff}), we see
that the term containing $l_6$ does vanish in the $q\to0$ limit.  The
cancellation at $q=0$ of the two summation terms from Eq.~(\ref{pionff}) is
easily demonstrated in the notation of a temporally infinite lattice,
\begin{eqnarray}
&&
\frac{1}{N_s^3}\sum_{\vec k}\int_{-\pi/a}^{\pi/a}\frac{dk_4}{2\pi}
 \left(2\sin^2ak_4D^2(k)-\cos ak_4D(k)\right) \nonumber \\
 &=& \frac{1}{N_s^3}\sum_{\vec k}\int_{-\pi/a}^{\pi/a}\frac{dk_4}{2\pi}
 \frac{d~}{dk_4}\left(-\sin ak_4D(k)\right) \nonumber \\
 &=& 0.
\end{eqnarray}

\begin{figure}[tb]
\includegraphics[width=10cm]{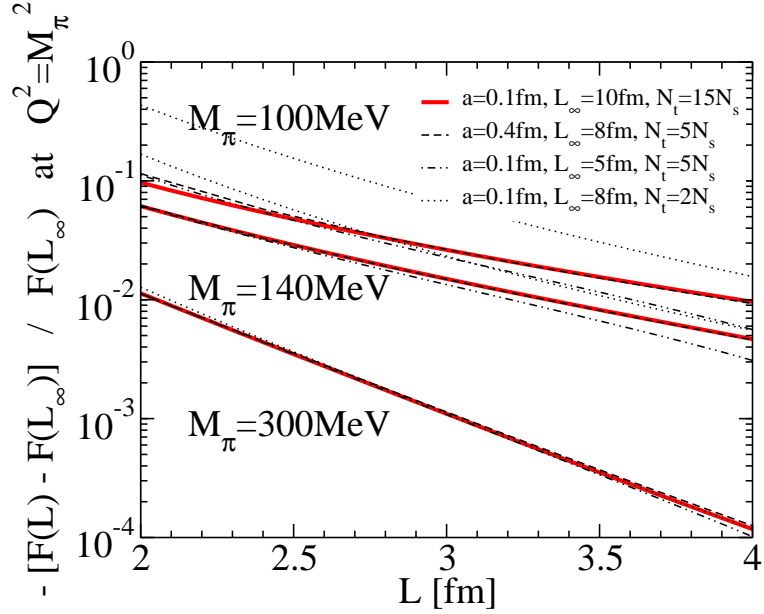}
\caption{Fractional change in the pion form factor at $Q^2=M_\pi^2$ as a
         function of spatial volume.}
\label{fig:pionffplot}
\end{figure}

Fig.~\ref{fig:pionffplot} shows the numerical results for the volume
dependence of the pion form factor at $q^2=M_\pi^2$ (meant to represent a
typical ChPT mass scale) obtained from Eq.~(\ref{pionff}),
where for numerical ease we work in the Breit frame.
As is evident from the plot, the form factor's fractional volume dependence
has a similar magnitude to that obtained for the pion mass and decay constant.

\begin{figure}[tb]
\includegraphics[width=10cm]{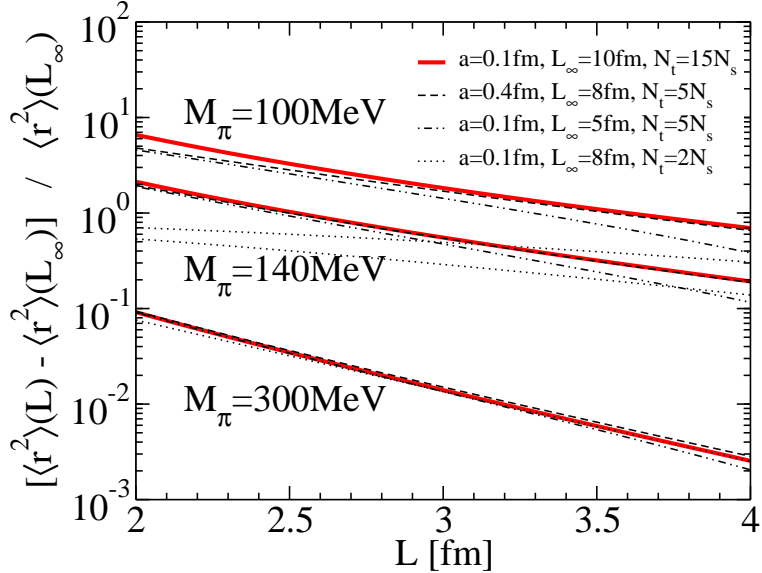}
\caption{Fractional change in the pion charge radius as a
         function of spatial volume.}
\label{fig:rsqplot}
\end{figure}

The pion charge radius is extracted from the slope of the form factor at
$q^2=0$.  Choosing $q=(0,0,q_3,0)$ for definiteness, we find 
\begin{eqnarray}
\left<r_\pi^2\right>_\pi &=& -6\lim_{q^2\to0}\frac{dF(q^2)}{dq^2}  \nonumber \\
 &=& \frac{-6l_6}{f^2} + \frac{12}{N_s^3N_tf^2}\sum_k\sin^2ak_4\left[
     \cos ak_3D^3(k)-4\sin^2ak_4\sin^2ak_3D^4(p)\right] \label{rsq}
\end{eqnarray}
A graph of the fractional volume dependence of this quantity is provided
in Fig.~\ref{fig:rsqplot}.  The magnitude of the effect is dramatically larger
than for the mass, decay constant and form factor simply because the
charge radius has (volume-dependent) loop contributions at its first nonzero
ChPT order.

\section{Summary and outlook}\label{sec:outlook}

The pion mass, decay constant, form factor and charge radius have been computed
from $O(p^4)$ chiral perturbation theory in a finite volume by using lattice
regularization.  A suggested advantage of this regularization scheme is that
the renormalization can be carried out numerically, leaving fewer analytical
steps to be performed.  Explicit expressions for these observables are given
as four-dimensional finite sums in Eqs.~(\ref{masseqn2}), (\ref{F}),
(\ref{pionff}) and (\ref{rsq}).

The dimensionally regularized expressions for the pion mass and decay constant
are known to be one-dimensional sums over Bessel
functions\cite{GLvolume1,ColHae}, and results from the two regularization
schemes agree numerically.  In essence, dimensional regularization arrives at a
more compact result (i.e. fewer summations) because in that method more of the
renormalization is done analytically.  Given the small computational cost of
the four-dimensional summations, the lattice regularized result is also quite
usable in practice.  In addition, Appendix~\ref{continuumlimit} demonstrates
that the continuum limit of the lattice regularized result is analytically
identical to the dimensional regularized expression, if one chooses to
complete the entire analytical calculation instead of the computational
scheme proposed in this work.

As emphasized in Ref.~\cite{Colangelo}, it is necessary to extend discussions
of volume dependence to the two-loop level, and perhaps beyond, so that the
rate of convergence can be explored.  In general, two-loop renormalization is
substantially more involved than the one-loop case so the reduction of
analytical effort obtained by using lattice regularization could be of
considerable value.  The extension of lattice regularization to two loops
will involve the determination of numerical values for the Lagrangian's low
energy constants (and their scale dependences) since they will no longer
subtract away in the difference between two volumes.  It will also require
an understanding of the interplay between power divergences, $1/a^n$, and
volume dependences, $1/(aN_s)^n$.  In particular, one does not want to rely
on numerical cancellations among diverging summations.
These issues are currently under investigation, in hopes of extending this
practical computational method to the domain of multi-loop ChPT calculations.
 
\acknowledgments

The authors thank Daniel Mazur for his involvement in the initial stages of
this research, Georg von Hippel for a careful reading of the manuscript,
and Hermann Krebs for useful discussions.
This work was supported in part by the Deutsche Forschungsgemeinschaft,
the Natural Sciences and Engineering Research Council of Canada,
and the Canada Research Chairs Program.

\begin{appendix}
\section{Lattice Feynman rules for the pion mass}\label{feynmanrules}

This appendix contains an explicit derivation of the Feynman rules required
for a one-loop computation of the volume dependence for the pion mass.
Only the third diagram of Fig.~\ref{fig:massfeyn} contributes, which
contains the pion propagator and the four pion vertex.

Beginning from Section~\ref{sec:Lagrangian} and using ellipses to denote
terms that do not contribute to the pion two-point function, the relevant
terms in the action are
\begin{eqnarray}
\delta S &=& a^4\sum_x{\cal L}_2(x) +\ldots \nonumber \\
         &=& -\frac{a^2f^2}{4}\sum_{x,\mu}{\rm Tr}\left[U^\dagger(x+\hat\mu)
             U(x)+U^\dagger(x)U(x+\hat\mu)\right] - \frac{a^4f^2}{2}Bm_q
             \sum_x{\rm Tr}\left[U(x)+U^\dagger(x)\right] \nonumber \\
          && + \ldots \nonumber \\
         &=& -a^2\delta^{ab}\sum_{x,\mu}\pi^a(x)[\pi^b(x+\hat\mu)
             -\pi^b(x)] + \frac{a^4x_\pi^2}{2}\delta^{ab}\sum_x\pi^a(x)\pi^b(x)
             + \ldots
\end{eqnarray}
with $\hat\mu = a \hat{e}_\mu$.
This will now be expressed in terms of the Fourier transform,
\begin{equation}
\pi(x) \equiv \frac{1}{N_s^3N_t}\sum_k\tilde\pi(k)e^{i k\cdot x},
\end{equation}
where the summation extends over the set of momenta
\begin{equation}
k = \left(\frac{2\pi n_1}{aN_s},\frac{2\pi n_2}{aN_s},\frac{2\pi n_3}{aN_s},
    \frac{2\pi n_4}{aN_t}\right)
\end{equation}
with
\begin{eqnarray}
n_j &=& -\frac{N_s}{2}, -\frac{N_s}{2}+1, \ldots, \frac{N_s}{2}-1, \\
n_4 &=& -\frac{N_t}{2}, -\frac{N_t}{2}+1, \ldots, \frac{N_t}{2}-1,
\end{eqnarray}
and we have chosen $N_s, N_t$ to be even, in order to keep the presentation
simple.
Using the relation
\begin{equation}
\sum_xe^{ik\cdot x} = N_s^3N_t\delta_{k,0}^{(4)}
\end{equation}
leads to
\begin{eqnarray}
\delta S &=& -\frac{a^2\delta^{ab}}{N_s^3N_t}\sum_{k,k^\prime,\mu}
             \delta_{k+k^\prime,0}^{(4)}\tilde\pi^a(k)\tilde\pi^b(k^\prime)
             \left[e^{ik^\prime\cdot\hat\mu}-1\right]
             +\frac{a^4x_\pi^2\delta^{ab}}{2N_s^3N_t}\sum_{k,k^\prime}
             \delta_{k+k^\prime,0}^{(4)}\tilde\pi^a(k)\tilde\pi^b(k^\prime)
             + \ldots \nonumber \\
 &=& \frac{a^4\delta^{ab}}{2N_s^3N_t}\sum_k\left[x_\pi^2+\frac{1}{2a^2}
             \sum_\mu(1-\cos k\cdot\hat\mu)\right]
             \tilde\pi^a(k)\tilde\pi^b(-k) + \ldots
\end{eqnarray}
The Euclidean two-point correlator for incoming pion fields $\tilde\pi^a(p)$
and $\tilde\pi^b(q)$ is
\begin{equation}
-\frac{d~}{d\left(\frac{\tilde\pi^a(p)}{N_s^3N_t}\right)}
 \frac{d~}{d\left(\frac{\tilde\pi^b(q)}{N_s^3N_t}\right)}
 \left(\frac{\delta S}{a^4N_s^3N_t}\right)
 = -\delta^{ab}\delta_{p+q,0}^{(4)}\left[x_\pi^2+\frac{2}{a^2}\sum_\mu
    (1-\cos p\cdot\hat\mu)\right].
\end{equation}
The pion propagator is the negative inverse of this expression with $p=-q$
and $a=b$, and is $a^2D(k)$ of Eq.~(\ref{piprop}).

To derive the four pion vertex, we again begin from
Section~\ref{sec:Lagrangian} and use ellipses to denote
terms that do not contribute,
\begin{eqnarray}
\delta S &=& a^4\sum_x{\cal L}_2(x) +\ldots \nonumber \\
         &=& -\frac{a^2f^2}{4}\sum_{x,\mu}{\rm Tr}\left[U^\dagger(x+\hat\mu)
             U(x)+U^\dagger(x)U(x+\hat\mu)\right] - \frac{a^4f^2}{2}Bm_q
             \sum_x{\rm Tr}\left[U(x)+U^\dagger(x)\right] \nonumber \\
          && + \ldots \nonumber \\
         &=& -\frac{a^2}{48f^2}{\rm Tr}(\tau^a\tau^b\tau^c\tau^d)\sum_{x,\mu}
     \left[\pi^a(x+\hat\mu)\pi^b(x+\hat\mu)\pi^c(x+\hat\mu)\pi^d(x+\hat\mu)
     \right.\nonumber \\
  && -4\pi^a(x+\hat\mu)\pi^b(x+\hat\mu)\pi^c(x+\hat\mu)\pi^d(x)
     +6\pi^a(x+\hat\mu)\pi^b(x+\hat\mu)\pi^c(x)\pi^d(x) \nonumber \\
  && \left.-4\pi^a(x+\hat\mu)\pi^b(x)\pi^c(x)\pi^d(x)
     +\pi^a(x)\pi^b(x)\pi^c(x)\pi^d(x)\right] \nonumber \\
  && -\frac{a^4x_\pi^2}{48f^2}{\rm Tr}(\tau^a\tau^b\tau^c\tau^d)
     \sum_x \pi^a(x)\pi^b(x)\pi^c(x)\pi^d(x)
     + \ldots \nonumber \\
  &=& -\frac{a^2}{48f^2N_s^{12}N_t^4}{\rm Tr}(\tau^a\tau^b\tau^c\tau^d)
      \sum_{x,\mu,k,k^\prime,k^{\prime\prime},k^{\prime\prime\prime}}
      \tilde\pi^a(k)\tilde\pi^b(k^\prime)\tilde\pi^c(k^{\prime\prime})
      \tilde\pi^d(k^{\prime\prime\prime})e^{i(k+k^\prime+k^{\prime\prime}
      +k^{\prime\prime\prime})\cdot x} \nonumber \\
   && \left[
      e^{i(k+k^\prime+k^{\prime\prime}+k^{\prime\prime\prime})\cdot\hat\mu}
      -4e^{i(k+k^\prime+k^{\prime\prime})\cdot\hat\mu}
      +6e^{i(k+k^\prime)\cdot\hat\mu}-4e^{ik\cdot\hat\mu}+1\right]
      \nonumber \\
   && -\frac{a^4x_\pi^2}{48f^2N_s^{12}N_t^4}{\rm Tr}(\tau^a\tau^b\tau^c\tau^d)
      \sum_{x,k,k^\prime,k^{\prime\prime},k^{\prime\prime\prime}}
      \tilde\pi^a(k)\tilde\pi^b(k^\prime)\tilde\pi^c(k^{\prime\prime})
      \tilde\pi^d(k^{\prime\prime\prime})e^{i(k+k^\prime+k^{\prime\prime}
      +k^{\prime\prime\prime})\cdot x} \nonumber \\
  &=& -\frac{a^2}{24f^2N_s^9N_t^3}{\rm Tr}(\tau^a\tau^b\tau^c\tau^d)
      \sum_{k,k^\prime,k^{\prime\prime}}
      \tilde\pi^a(k)\tilde\pi^b(k^\prime)\tilde\pi^c(k^{\prime\prime})
      \tilde\pi^d(-k-k^\prime-k^{\prime\prime}) \nonumber \\
   && \left[2a^2x_\pi^2+\sum_\mu
      \left(1-2e^{ik\cdot\hat\mu}-3e^{i(k+k^\prime)\cdot\hat\mu}
      -2e^{i(k+k^\prime+k^{\prime\prime})\cdot\hat\mu}\right)\right].
\end{eqnarray}
The Euclidean four-point correlator (i.e. the Feynman rule)
for incoming pion fields $\tilde\pi^a(p)$,
$\tilde\pi^b(q)$, $\tilde\pi^c(r)$ and $\tilde\pi^d(-p-q-r)$ is
\begin{eqnarray}
&&
-\frac{d~}{d\left(\frac{\tilde\pi^a(p)}{N_s^3N_t}\right)}
 \frac{d~}{d\left(\frac{\tilde\pi^b(q)}{N_s^3N_t}\right)}
 \frac{d~}{d\left(\frac{\tilde\pi^c(r)}{N_s^3N_t}\right)}
 \frac{d~}{d\left(\frac{\tilde\pi^d(-p-q-r)}{N_s^3N_t}\right)}
 \left(\frac{\delta S}{a^4N_s^3N_t}\right) \nonumber \\
 &=& \frac{2}{3a^2f^2}\left(\delta^{ab}\delta^{cd}+\delta^{ac}\delta^{bd}
   +\delta^{ad}\delta^{bc}\right)\sum_\mu\left(1-\cos p_\mu-\cos q_\mu
   -\cos r_\mu-\cos (p+q+r)_\mu\right) \nonumber \\
&& +\frac{x_\pi^2}{3f^2}\left(\delta^{ab}\delta^{cd}+\delta^{ac}\delta^{bd}
   +\delta^{ad}\delta^{bc}\right)
   +\frac{2}{a^2f^2}\delta^{ab}\delta^{cd}\sum_\mu\cos (p+q)_\mu
   \nonumber \\
&& +\frac{2}{a^2f^2}\delta^{ac}\delta^{bd}\sum_\mu\cos (p+r)_\mu
   +\frac{2}{a^2f^2}\delta^{ad}\delta^{bc}\sum_\mu\cos (q+r)_\mu.
\end{eqnarray}
The $a\to0$ limit reproduces the standard continuum Feynman rule as expected.

\section{Analytic derivation of the continuum limit}\label{continuumlimit}

In the main body of this article, lattice regularized results were left in the
form of
loop summations over products of Feynman rules, since this is sufficient
to produce numerical results.  As expected, the numerics agreed
with analytic dimensional regularized calculations where available, since
physics does not depend on regularization scheme.
Using the volume dependence of the pion mass as an explicit example, this
appendix verifies that continuing
the analytic steps in the lattice regularization approach, and taking
the continuum limit, leads to the same analytic result that comes from
dimensional regularization.

{}From Eqs.~(\ref{masseqn1}) and (\ref{masseqn2}), lattice regularization
produces the following difference in pion mass for two lattice volumes,
\begin{eqnarray}
\left[M_\pi(L)-M_\pi(L^\prime)\right]_{{\rm finite~}a} &=& \lim_{N_t\to\infty}
      \left[\frac{x_\pi}{4a^2f^2N_s^3N_t}
      \sum_k\frac{3-2\cos ak_4}{a^2x_\pi^2+2\sum_\mu(1-\cos ak_\mu)}\right.
      \nonumber \\
    && \left.-\frac{x_\pi}{4a^2f^2(N_s^\prime)^3N_t}
      \sum_{k^\prime}\frac{3-2\cos ak_4}{a^2x_\pi^2+2\sum_\mu(1-\cos a
      k_\mu^\prime)}\right] + O(a), \quad \label{massdiff}
\end{eqnarray}
where the summations extend over the set of momenta
\begin{equation}
k = \left(\frac{2\pi n_1}{aN_s},\frac{2\pi n_2}{aN_s},\frac{2\pi n_3}{aN_s},
    \frac{2\pi n_4}{aN_t}\right),~~~
k^\prime = \left(\frac{2\pi n_1^\prime}{aN_s^\prime},\frac{2\pi n_2^\prime}
           {aN_s^\prime},\frac{2\pi n_3^\prime}{aN_s^\prime},
           \frac{2\pi n_4}{aN_t}\right),
\end{equation}
with
\begin{eqnarray}
n_j &=& -\frac{N_s}{2}, -\frac{N_s}{2}+1, \ldots, \frac{N_s}{2}-1, \\
n_j^\prime &=& -\frac{N_s^\prime}{2}, -\frac{N_s^\prime}{2}+1, \ldots,
               \frac{N_s^\prime}{2}-1, \\
n_4 &=& -\frac{N_t}{2}, -\frac{N_t}{2}+1, \ldots, \frac{N_t}{2}-1
\end{eqnarray}
and $N_s, N'_s, N_t, N'_t$ even.

Using a double angle formula from basic trigonometry,
Eq.~(\ref{massdiff}) can be re-expressed as
\begin{eqnarray}
\left[M_\pi(L)-M_\pi(L^\prime)\right]_{{\rm finite~}a}
 &=& \lim_{N_t\to\infty}\left[\frac{x_\pi}
      {4a^2f^2N_s^3N_t}
      \sum_k\frac{1+4\sin^2(ak_4/2)}{a^2x_\pi^2+4\sum_\mu\sin^2(ak_\mu/2)}
      \right. \nonumber \\
    && \left.-\frac{x_\pi}{4a^2f^2(N_s^\prime)^3N_t}
      \sum_{k^\prime}\frac{1+4\sin^2(ak_4^\prime/2)}{a^2x_\pi^2+4\sum_\mu\sin^2
      (ak_\mu^\prime/2)}\right] + O(a).
\end{eqnarray}
Since this mass difference is finite even in the continuum limit, and since
the Taylor expansion of $\sin\theta$ satisfies absolute convergence
term by term over the entire range of interest, $-\pi/2\leq\theta\leq\pi/2$,
the leading volume dependence is obtained by retaining the leading term in
this Taylor expansion.  The result is
\begin{eqnarray}
\left[M_\pi(L)-M_\pi(L^\prime)\right]_{{\rm finite~}a}
&=& \frac{x_\pi}{4a^4f^2}\lim_{N_t\to\infty}
      \left[\frac{1}{N_s^3N_t}\sum_k\frac{1}{x_\pi^2+\sum_\mu k_\mu^2}
      -\frac{1}{(N_s^\prime)^3N_t}
      \sum_{k^\prime}\frac{1}{x_\pi^2+\sum_\mu k_\mu^\prime}\right]
    \nonumber \\
 && + O(a).
\end{eqnarray}
As stated in Sec.~\ref{sec:mass}, our goal is to
consider the dependence of observables on spatial volume in the double
limit $a\to0$, $T\equiv aN_t\to\infty$ with $L\equiv aN_s$ (and
for now $L^\prime\equiv aN_s^\prime$ also) held fixed.  With this simple
change of variables (but leaving $a$ finite momentarily), we obtain
\begin{eqnarray}
\left[M_\pi(L)-M_\pi(L^\prime)\right]_{{\rm finite~}a}
 &=& \frac{x_\pi}{4f^2}\lim_{T\to\infty}
      \left[\frac{1}{L^3T}\sum_n\frac{1}{x_\pi^2+\sum_j(2\pi n_j/L)^2
      +(2\pi n_4/T)^2} \right. \nonumber \\
 &&   \left.-\frac{1}{(L^\prime)^3T}
      \sum_{n^\prime}\frac{1}{x_\pi^2+\sum_j(2\pi n_j^\prime/L^\prime)^2
      +(2\pi n_4/T)^2}\right] + O(a) \nonumber \\
 &=& \frac{x_\pi}{4f^2}\int_{-\pi/a}^{\pi/a}\frac{dp_4}{2\pi}
      \left[\frac{1}{L^3}\sum_{\vec n}\frac{1}{x_\pi^2+\sum_j(2\pi n_j/L)^2
      +p_4^2} \right. \nonumber \\
 &&   \left.-\frac{1}{(L^\prime)^3}
      \sum_{\vec n^\prime}\frac{1}{x_\pi^2+\sum_j(2\pi n_j^\prime/L^\prime)^2
      +p_4^2}\right] + O(a).
\end{eqnarray}
We now take the continuum limit, which
merely extends the bounds of summation and integration to
$\pm\infty$.  The integral over $p_4$ can be performed in closed form,
\begin{eqnarray}
M_\pi(L)-M_\pi(L^\prime)
&=&\frac{x_\pi}{8f^2}
   \left[\frac{1}{L^3}\sum_{\vec n}\frac{1}{\sqrt{x_\pi^2+\sum_j(2\pi n_j/L)^2}}
   -\frac{1}{(L^\prime)^3}
   \sum_{\vec n^\prime}\frac{1}{\sqrt{x_\pi^2+\sum_j(2\pi n_j^\prime/
   L^\prime)^2}}\right] \nonumber \\
&=&\frac{x_\pi}{8f^2\sqrt{\pi}}\int_0^\infty{ds}\frac{e^{-sx_\pi^2}}{\sqrt{s}}
   \left[\frac{1}{L^3}\sum_{\vec n}e^{-s\sum_j(2\pi n_j/L)^2}
   -\frac{1}{(L^\prime)^3}\sum_{\vec n^\prime}e^{-s\sum_j(2\pi n_j^\prime/
   L^\prime)^2}\right]. \nonumber \\
\end{eqnarray}
Letting $L^\prime\to\infty$ gives
\begin{eqnarray}
M_\pi(L)-M_\pi
&=&\frac{x_\pi}{8f^2\sqrt{\pi}}\int_0^\infty{ds}\frac{e^{-sx_\pi^2}}{\sqrt{s}}
   \left[\frac{1}{L^3}\sum_{\vec n}e^{-s\sum_j(2\pi n_j/L)^2}
   -\int_{-\infty}^\infty\frac{d^3p}{(2\pi)^3}e^{-s\vec p^2}\right]
   \nonumber \\
&=&\frac{x_\pi}{8f^2\sqrt{\pi}}\int_0^\infty{ds}\frac{e^{-sx_\pi^2}}{\sqrt{s}}
   \left[\frac{1}{L^3}\sum_{\vec n}e^{-s\sum_j(2\pi n_j/L)^2}
   -\frac{1}{(4\pi s)^{3/2}}\right],
\end{eqnarray}
where $M_\pi\equiv M_\pi(\infty)$.
We can now make use of a relation that appears in Ref.~\cite{BecVil},
\begin{equation}
\sum_{n=-\infty}^\infty e^{-\tau n^2}
 = \sqrt{\frac{\pi}{\tau}}\sum_{n=-\infty}^\infty e^{-\pi^2n^2/\tau},
\end{equation}
to obtain
\begin{eqnarray}
M_\pi(L)-M_\pi
&=&\frac{x_\pi}{64f^2\pi^2}\int_0^\infty{ds}\frac{e^{-sx_\pi^2}}{s^2}
   \left[\sum_{\vec n}e^{-\vec n^2L^2/(4s)}-1\right] \nonumber \\
&=&\frac{x_\pi}{64f^2\pi^2}\int_0^\infty{ds}\frac{e^{-sx_\pi^2}}{s^2}
   \sum_{\vec n\neq\vec 0}e^{-\vec n^2L^2/(4s)} \nonumber \\
&=&\frac{x_\pi^2}{16f^2\pi^2L}\sum_{\vec n\neq\vec 0}
   \frac{K_1(x_\pi L\sqrt{n_1^2+n_2^2+n_3^2})}{\sqrt{n_1^2+n_2^2+n_3^2}},
\end{eqnarray}
where $K_1(x)$ is a Bessel function of the second kind.
This is the result known from dimensional regularized
calculations.\cite{GLvolume1,ColDur}
Though expressed as a triple summation over $n_1$, $n_2$ and $n_3$, the
function only contains the sum of squares, $n_1^2+n_2^2+n_3^2$, thus
allowing $M_\pi(L)-M_\pi$ to be
represented by a one-dimensional summation when multiplicity factors are
defined.\cite{ColDur}

\end{appendix}

\end{document}